# Twin GEM-TPC Prototype (HGB4) Beam Test at GSI and Jyväskylä – a Development for the Super-FRS at FAIR

F. García[*1], R. Turpeinen[1], J. Äystö[1,2], T. Grahn[2], S. Rinta-Antila[2], A. Jokinen[2], J. Kunkel[3], V. Kleipa[3], A. Gromliuk[3], H. Risch[3], C. Caesar[3], C. Simons[3], C. J. Schmidt[3], A. Prochazka[3], J. Hoffmann[4], I. Rusanov[4], N. Kurz[4], H. Heggen[4], P. Strmen[5], M. Pikna[5], B. Sitar[5]

[1]Helsinki Institute of Physics, University of Helsinki, 00014 Helsinki, Finland
[2]Department of Physics, University of Jyväskylä, Jyväskylä, Finland
[3]Detector Laboratory, GSI Helmholtzzentrum, Darmstadt 64291, Germany
[4]Experimental Electronics Department, GSI Helmholtzzentrum, Darmstadt 64291, Germany
[5]FMFI Bratislava, Comenius University, Bratislava, Slovakia

*Abstract*–The FAIR[1] facility is an international accelerator centre for research with ion and antiproton beams. It is being built at Darmstadt, Germany as an extension to the current GSI research institute. One major part of the facility will be the Super-FRS[2] separator, which will be include in phase one of the project construction. The NUSTAR experiments will benefit from the Super-FRS, which will deliver an unprecedented range of radioactive ion beams (RIB). These experiments will use beams of different energies and characteristics in three different branches; the high-energy which utilizes the RIB at relativistic energies 300-1500 MeV/u as created in the production process, the low-energy branch aims to use beams in the range of 0-150 MeV/u whereas the ring branch will cool and store beams in the NESR ring. The main tasks for the Super-FRS beam diagnostics chambers will be for the set up and adjustment of the separator as well as to provide tracking and event-by-event particle identification. The Helsinki Institute of Physics, and the Detector Laboratory and Experimental Electronics at GSI are in a joint R&D of a GEM-TPC detector which could satisfy the requirements of such tracking detectors, in terms of tracking efficiency, space resolution, count rate capability and momenta resolution. The current prototype, which is the generation four of this type, is two GEM-TPCs in twin configuration inside the same vessel. This means that one of the GEM-TPC is flipped on the middle plane w.r.t. the other one. This chamber was tested at Jyväskylä accelerator with protons projectiles and at GSI with Uranium, fragments and Carbon beams during this year 2016.

## I. Introduction

THE FAIR facility at GSI will stand for the research with ion and antiproton beams, in cooperation with their users and the international community. With the new project, GSI aims to provide scientists in Europe and the world with an outstanding accelerator and experimental facility for studying matter at the level of atoms, atomic nuclei, protons and neutrons as the building blocks of nuclei - and part of a wider family called hadrons - and the subnuclear constituents called quarks and gluons.

The facility will provide an extensive range of beams from protons and antiprotons to ions up to uranium with world record intensities and excellent beam quality in the longitudinal as well as transverse phase space. The scientific goals pursued at FAIR[3] include:

• Studies with beams of short-lived radioactive nuclei, aimed at revealing the properties of exotic nuclei, understanding the nuclear properties that determine what happens in explosive processes in stars and how the elements are created, and testing fundamental symmetries.

• The study of hadronic matter at the subnuclear level with beams of anti-protons, in particular of the following key aspects: the confinement of quarks in hadrons, the generation of hadron masses by spontaneous breaking of chiral symmetry, the origin of the spins of nucleons, and the search for exotic hadrons such as charmed hybrid mesons and glueballs.

• The study of compressed, dense hadronic matter through nucleus-nucleus collisions at high energies.

Manuscript received December 13, 2016. This work was supported by the Ministry of Education of Finland.

F. García* is with the Helsinki Institute of Physics, University of Helsinki, P.O. Box 64, FI-00014 University of Helsinki, Finland (telephone: +358-9-19151086, e-mail: Francisco.Garcia@helsinki.fi).

R. Turpeinen, J. Äystö, are with the Helsinki Institute of Physics, University of Helsinki, P.O. Box 64, FI-00014 University of Helsinki, Finland (e-mails: Raimo.Turpeinen@helsinki.fi, Juha.Aysto@helsinki.fi, ).

T. Grahn, S. Rinta-Antila, A. Jokinen are with the Department of Physics, University of Jyväskylä, Jyväskylä, Finland (e-mails: tuomas.grahn@jyu.fi, sami.rinta-antila@jyu.fi, ari.jokinen@jyu.fi)

B. Voss, J. Kunkel, V. Kleipa, A. Gromliuk, H. Risch, C. Caesar, C. Simons, C. J. Schmidt, A. Prochazka are with the Detector Laboratory at GSI, 64291 Darmstadt, Germany (e-mails: B.Voss@gsi.de, j.kunkel@gsi.de, v.kleipa@gsi.de, Andrii Gromliuk agromliuk@gmail.com, H.Risch@gsi.de, C.Caesar@gsi.de, C.Simons@gsi.de, C.J.Schmidt@gsi.de, a.prochazka@gsi.de).

J. Hoffmann, I. Rusanov, N. Kurz, H. Heggen are with the Experimental Electronics Department at GSI, 64291 Darmstadt, Germany (e-mails: j.hoffmann@gsi.de, I.Rusanov@gsi.de, N.Kurz@gsi.de, H.Heggen@gsi.de).

M. Pikna, B. Sitar, P. Strmen are with the FMFI Bratislava, Comenius University, Bratislava, 84248 Slovakia (e-mails: pikna@fmph.uniba.sk, sitar@fmph.uniba.sk, strmen@fmph.uniba.sk).

• The study of bulk matter in the high density plasma state, a state of matter of interest for inertial confinement fusion and for various astrophysical sites.
• Studies of Quantum electrodynamics (QED), of extremely strong electromagnetic.

The NuSTAR experiments will be dedicated to the study of Nuclear Structure, Astrophysics and Reactions. In particular with the use of beams of radioactive species separated and identified by the Superconducting Fragment Recoil Separator (Super-FRS). No experimental programme in nuclear physics stands entirely on its own and NuSTAR is not an exception. Strictly speaking some activities at GSI will fall naturally into the sphere of interest of NuSTAR, such as the search for new superheavy elements and the study of their physical and chemical properties. Since the success of the experimental programme of NuSTAR depends critically on the specifications and resulting properties of the Super-FRS, this device will be also seen as an integral part of its activities.

The fragment separator FRS provides high energy, spatially separated monoisotopic beams of exotic nuclei of all elements up to uranium[4,5]. The fragments are separated in flight, thus the accessible lifetimes are determined by the time-of-flight through the ion-optical system, which range from the submicrosecond domain upwards.

The FRS[4] has proven to be an extremely versatile instrument for new nuclear and atomic studies, as well as for experiments in applied physics. More than 150 new isotopes have been discovered and studied directly at the FRS, including the doubly magic nuclei $^{100}$Sn and $^{78}$Ni. Neutron-deficient projectile fragments between lead and uranium have been excited by the Coulomb field of heavy targets to investigate low-energy fission properties for a large number of nuclei for the first time.

Additionally, the FRS delivers fragment beams for decay and reaction studies to the ALADIN-LAND setup and the KAOS spectrometer. Fragments are also delivered to the ESR, for experiments such as precision mass and lifetime measurements. Moreover, the FRS has been used as a high-resolution magnetic spectrometer. Precise momentum measurements led to the discovery of new halo properties, the first observation of deeply bound pionic states in heavy nuclei, and several fundamental properties in atomic collision physics. In several fields of applied physics important studies were performed, such as the full characterization of heavy-ion therapy beams and optimization of PET diagnostics for the GSI cancer therapy project.

Although the FRS facility has contributed to the field of heavy-ion science with great success, there are four major enhancements that can be applied to improve the method considerably: 1. Higher intensity of the primary beams, 2. increased transmission for fission fragments produced by uranium projectiles, 3. improved transmission of fragments to the dedicated experimental areas, 4. larger acceptance of fragments by the storage cooler ring. These improvements are addressed in the recently proposed international rare-isotope facility at GSI, which includes a powerful driver accelerator, a large-acceptance superconducting in-flight separator, and a storage-cooler ring complex optimized to accept large-emittance secondary beams.

The Superconducting fragment Separator (Super-FRS)[5] is a powerful in-flight facility which will provide spatially separated isotopic beams up to elements of the heaviest projectiles. It is superior to the present FRS due to the incorporation of more separation stages and larger magnet apertures through the use of superconducting coils. The Super-FSR is based on results, experience, methods and techniques which were pioneered and developed for relativistic heavy ions at the FRS.

## II. SUPER FRS DIAGNOSTIC CHAMBERS

It is planned to implement a detection system that can be commonly used for all experiments at the different Super-FRS branches and comes with its associated data acquisition scheme. The main task of this combined system is threefold:
- it can be used to set up and adjust the separator,
- it provides the necessary measures for machine safety and monitoring,
- it allows for an event-by-event particle identification, tracking and characterization of the produced rare ion species.

Furthermore, the beam intensities at different locations in the separator are to be monitored, e.g. as to normalize measured rates to be able to extract absolute cross sections. The modi operandi depend strongly on these given tasks and the necessary requirements for the combined detector and acquisition systems will be given.

Setting up and adjusting the separator can be done at a low rate for almost any detector system. The main design goal is to get an easy to maintain, reliable system. An online monitoring has to be performed, especially in the target and beam catcher areas. Any deviation of the primary beam from its nominal position should lead immediately to an interlock condition. The main challenge is to cope with the very high intensities and background radiation here. The design of the detector systems should allow extended periods of operation without maintenance hands-on.

For almost all experiments, the separator is to be treated as first part of the experimental setup. The beam particles entering the different branches have to be identified and their longitudinal and transverse momentum components should be known. For tracking experiments to be carried out in the Low-Energy Branch and the High-Energy Branch, the measurement has to be performed on an event-by-event basis. This implies that the data acquisition system of the particular experiments and the Super-FRS should be closely coupled if not identical. The Super-FRS data taking will therefore be designed in accordance with the common NUSTAR data acquisition scheme. The requirements on detector systems are demanding at the entrance of the main separator, where rates up to $10^9$ particles /s can be expected.

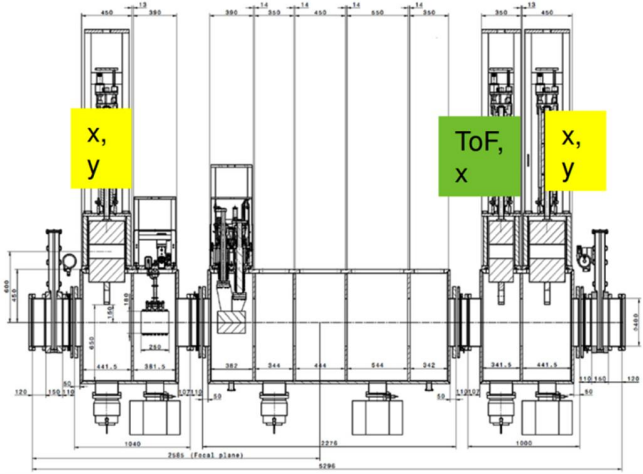

Fig. 1. Super-FRS beam diagnostics chamber at the Main separator (FMF2).

The locations for the different detection systems are shown in Figure 1. It is foreseen to be able to run for about one year without opening the sections along the S-FRS. Generally we foresee UHV material although a pressure of $10^{-7}$ mbar is sufficient. The choice of the particular detector systems is driven by the idea trying to benefit from the various developments that are currently done in the detector laboratory and accelerator division.

### III. GEM-TPC DETECTOR DEVELOPMENT

The Diagnostics for slowly extracted beams are defined as slow extraction when the extraction times are above 100 ms. For this purpose a twin Time Projection Chambers with Gas Electron Multipliers (HGB4) will be installed and the requirements for such detectors are:
- No interference with the beam
- Large dynamic range
- Intensities of the order of 100 kHz

In order to satisfy these requirements a working group was formed, which produce several prototypes[6,7,8]. These chambers were commissioned in the laboratory and then tested in real beam conditions in the FRS at GSI.

This particular prototype HGB4 (Helsinki-GSI-Bratislava prototype 4), is the fourth generation of this type of GEM-TPC and the difference with previous ones is basically the introduction of the twin configuration, which mean that in one vessel there are two GEM-TPCs inside located side by side and one is flipped on the midplane against the other one. This configuration will form a tracking detector which has two field cages with opposite electric fields; therefore for a single track the drift distance traversed by the electrons will be different. However the sum of both drift times of those electrons has to be a constant number, thus bellowing to a single track. This method allows us to drastically reduce the ambiguity of association of hits to a track and keep the tracking efficiency close to 100% at high rate.

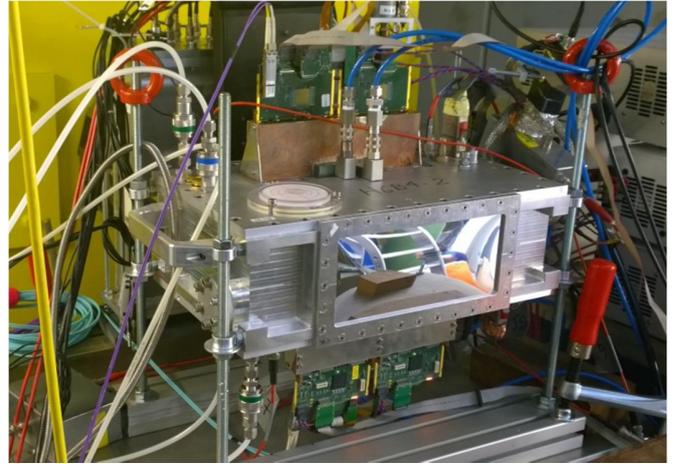

Fig.2. Super-FRS GEM-TPC prototype HGB4, equipped with four GMX-NYXOR cards.

The readout board has a geometry defined by 512 strips, which are 250 μm in width and pitch of 400 μm. This type of readout geometry suits very well to cope with high count rate and low dynamic range of the readout electronics. In the Fig. 3 is shown a close up of the readout strips.

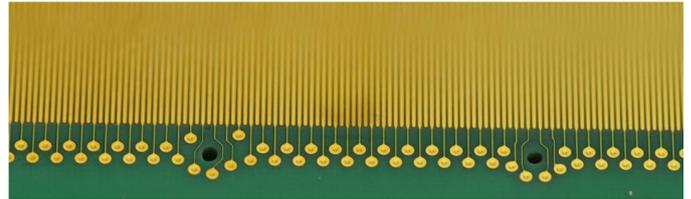

Fig. 3. Readout Pad plane consists of 512 strips, each of them has a width of 250 μm and a pitch of 400 μm.

During previous campaigns[9] we were able to test different readout systems and this year in particular the test were performed with the GMX-NYXOR cards, which contains two n-xyters[10] chips to readout a total of 256 channels per card.

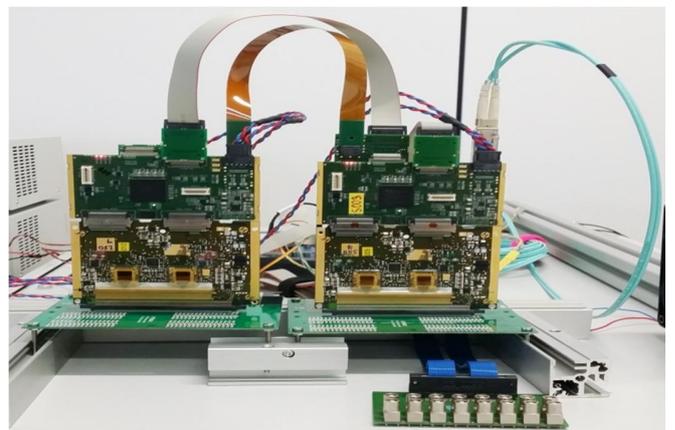

Fig. 4. The readout system GMX-NYXOR cards during testing in the laboratory.

## IV. RESULTS FROM TEST BEAM AT JYVÄSKYLÄ AND FRS

During this year we have participating in several test beam campaigns among of them one at Jyväskylä accelerator and two at the FRS at GSI.

The geometry at Jyväskylä test beam was quite simple and consists of a trigger scintillator and the HGB4 prototype, which means we could not have a reference tracker because the primary energy was quite low. The beam characteristics were primary projectiles protons with energy of 50 MeV/u and beam spot of about 50 cm at the location of the trigger scintillator, which was located at the entrance window of the HGB4 chamber.

However we could perform several scans and the most important of them was the intensity scan; were we could run the chamber from 1 kHz up to 7.78 MHz. This runs allow us to test the operation stability in high rate conditions, moreover the gain of the GEM stack was high, which added to the test the possibility to stress at the maximum the GEM stack of each of the GEM-TPC.

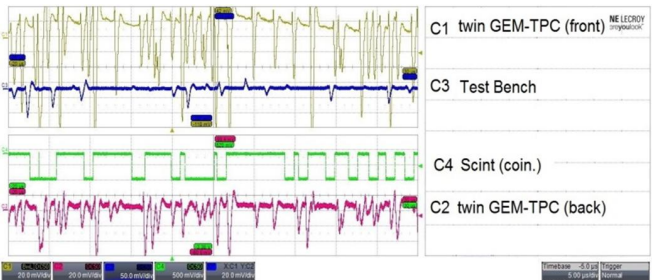

Fig. 5. The HGB4 signals taken from the Bottom of both GEM #3 at 2.20 MHz rate under Protons irradiation at Jyväskylä accelerator.

The second campaign was at FRS with primary beam particles of $^{238}$U with energy of 550 MeV/u and the intensity was varying up to $10^7$ ions/spill. The spills has a duration of about 0.5 - 10 s and the beam was moved from the center to left to the right, in addition to that the beam was focused and defocused.

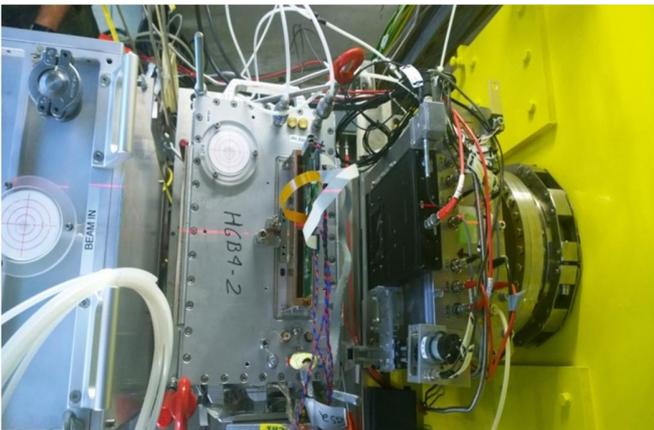

Fig. 6. The prototype HGB4 (chamber 2) located at S4 in the FRS, in between the reference tracker.

Below is shown a correlation plot between the two GEM-TPCs of the HGB4 chamber. This allows us to get the position of the beam and at the same time check the tracking performance of each of the GEM-TPCs (See Fig.7). One can see that the hits are concentrated on around 320 ch for both GEM-TPCs which means slightly shifted from the center in X coordinate.

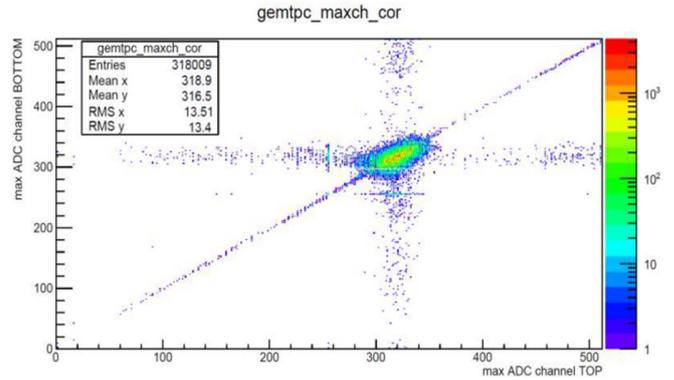

Fig. 7. Position correlation of Top and Bottom GEM-TPCs of the HGB4 prototype for Uranium projectiles.

In the Fig. 8 is shown single clusters produced by the Uranium beam. Each of them represents a single track traversing the HGB4 chamber.

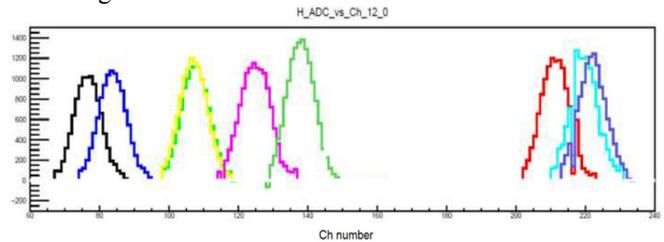

Fig. 8. Single clusters for Uranium beam.

These single clusters are in complete agreement with what has been simulating in Garfield for the Drift length of 10cm in P10 gas, taking into account the contribution given by the GEM stack in terms of diffusion. In addition to that the readout pad plane geometry was included in the simulation as well. This information was crucial to understand that the chamber was operating correctly and that the induced signals corresponded to the hits produced by a track and no instabilities in the operation of the field cages were observed.

## V. CONCLUSIONS

Results from the test Beam campaign at Jyväskylä and GSI has shown that twin GEM-TPC (HGB4) has operated very stable in continuous mode and the concept has been proven to be the final one. The next step is to analyze all the data collected to be able to obtain the tracking efficiency versus rate and its position resolution. We are confident that the detector efficiency was close to 100% for all of the runs and this will presented in the near future.

ACKNOWLEDGMENT

We will want to thank to the Finnish Ministry of Education for the long term funding of the Fair facility participation. Many thanks for the Jyväskylä accelerator team and in particular to the RADEF cave personal as well. We want to extend our gratitude to the FRS team for providing full support during the whole campaign.